# Transition of University to Prosumer Consortium Energy Model

Min Set Aung [2], Kaung Si Thu [1], Weerakorn Ongsakul [1], and Nimal Madhu Manjiparambil [1,*]

[1] Department of Energy, Environment and Climate Change,
School of Environment Resources and Development (SERD)
Asian Institute of Technology, Pathum Thani, 12120, Thailand

[2] Department of Information and Communication Technologies,
School of Engineering and Technology (SET)
Asian Institute of Technology, Pathum Thani, 12120, Thailand

*Corresponding author: mm-nimal@ait.ac.th

*Abstract* — The conventional electricity grid system has many critical issues, which could lead to power instability, blackout, and many other issues. One main perspective is to consume the relative amount of distributed renewable energy in the adaptation system. Prosumer Consortium Model in a small community can be one of the solutions to reduce the price of electricity, increase efficiency and provide a better electronification for University, moreover, to solve the frequent blackout of electricity at university level in developing countries. It can be the learning opportunity for the researchers as well as the complementary benefit for the consumers inside university's grid.

I. INTRODUCTION

The increased of greenhouse gas emission and constantly high demand for energy encourage the mechanism of electricity generation to increase the production of green energy and high efficiency. The factors mainly cause the usage of renewable energy production with distributed generations (DG) through the macro/micro power grid. The current electricity system does not have the ability to maintain with DG as the main energy source, thus, only a backup or supplementary source. The key to play the prosumer consortium market model in the grid system is the support of distributed energy resources as in macro or microgrid, which can bring the energy balance in the grid. This implementation of prosumer consortium energy model allows the benefit to the local customers which allows from arranging dispatchable load at peak power consumption, reduce the load on distribution and transmission system.

The model is being conferred in the experimental studies in which many pilot projects are running in communities, nonetheless, an instant implementation can cause the current system to a revolution of operation, control and storage technology. The market is established on the supply and demand of the electricity especially volatile during peak hours. The comprising features of a university can be stated as a community grid. The model grid promotes the incorporation at the level of consumption of renewable energy sources (RES) and energy storage systems (ESS) targeting to increase power quality, reliability and performance.

II. CHARACTERISTIC OF TRADING MODEL

1. Integration of ICT in the market model

The characteristic of the model has to be changed from a unidirectional line to a bidirectional power exchange, as well as ICT infrastructure to transfer the data between customer nodes. The prosumer consortium model comprises DERs such as micro-wind generator, solar generator, biomass, EV station, different types of ESS and fuel cells. The suitable preparation of the model has to be reliable and stable. When it comes to loads from a different consumer, they can be mainly distinguished into controllable load and uncontrollable load. The controllable load defines that the consumption of electricity can be stopped or start the operation in a matter of anticipated time, while incontrollable load does not have the flexible measure or control to operate in the time. Thus, these mentioned loads have to be aware of planning the demand response of the prosumer model. Sophisticated systems and advanced networking systems will have to the main part of transfer consumption/generation data, users' information and credit transfer. The aspects of the information system observe as software to approach the consumption information of building or node through web service as a worldwide web system to interact with the central monitoring of the prosumer consortium model. Then the web service can interact with the hardware interaction of the model, whether from the side of the consumer or supervisory unit. Communication of all the service intends to support the required and adequate bandwidth of the network as in an orthodox and reliable broadband power line technology. However, the convergence of 5G communication will be the future implication for the grid system. The complex control of the system is provided by the utilization of microprocessors to operate the load controllers, inverters and other components of the system.

2. Impact of Renewable Energy Production in University

The objective of energy generation is to shift from a conventional national grid system to take advantage of renewable energy production. However, the location of university and the available area, in order to obtain the required resources for RE production, depending on the method of power production such as biomass or wind energy. Those systems can be considered as a limitation to produce energy. Therefore, the popular source of supplying the required energy of the university can be powered by photovoltaic panels and microturbine, and besides, injections of power from electric vehicle can be an additional source of supporting RE. One of the important parts to maintain the RE power generation and local loads is a storage system – energy storage devices. The aggregated power from the surplus generation of RE will be solved by supplying back to the local demand or supporting the fulfill the local grid quality by discharging the storage devices. However, there is a thing that the implementation of energy storage system by itself cannot play as a vital part of the prosumer consortium model, thus the model can be designed without the storage. Since there are varieties of energy storage systems depending on the materials and energy, the effective system must be considered not only for financial assistance but the significant energy storage system to run the model. The application of energy storage system is mainly categorized into renewable energy integration, customer energy management, ancillary service and bulk energy. Typically, two configurations of energy storage system can be found in many systems, distributed and aggregated energy storage system. Generation output depends upon the cell temperature $T$ and solar radiance $G$, nevertheless, the efficiency and power loss of the solar panel has to be considered. Normally, the solar generation inclines gradually from the morning and reaches the peak production at noon and falls back in the afternoon like a bell curve. A sudden rise in energy consumption of local load must be acutely considered that the consumption level reaches at less time. The microturbine can be another solution to provide the prosumer consortium model with a significant role as supplementary.

3. Prosumer Consortium in the system

A prosumer consortium microgrid is most likely to be found in regions with high retail electricity price or high MS financial support levels and both conditions are very likely to occur simultaneously. Individual or several prosumer bodies will own microgenerations and operate DERs to obtain profits from the integration. Individual or several prosumer bodies will own micro/macro sources and operate DERs to obtain profits from the integration. Microsource owners will get revenue from the sale and minimize the bill. The prosumer consortium market model for a small community as university might play a role since the model tends to have smaller DERs and discrete storage system. It will try to be more independent on the main grid and take no notice of network constraints. Figure 1 shows the prosumer consortium market model for a small community grid with the internal financial flow, which does not trade with wholesale market and DSO, generally, the system works.

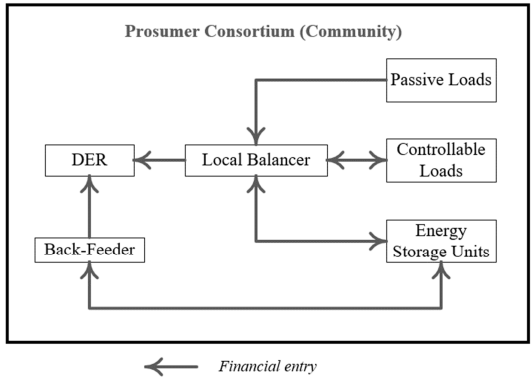

Figure 1 Financial entry in local prosumer consortium model.

III. FEASIBLE SYSTEM FOR THE MARKET MODEL

The form of DER and size of the generation mainly depends on the size of a university, funding and available resources to provide microgenerations. Meanwhile, a simple communication system with a small size of renewable energy generations can be premeditated. The installation of AMI at every consumer and prosumer node is necessary for data recording and operation of energy trade. The power line communication (PLC) broadband will be used while connecting to the central controller and smart meter, in which it will store the consumption/generation data and event statistics. The ability of BPL modems is to amplify the transmitted and received signal and perform the power converters gateway. The generation data of every generation units will be monitored by a central controller and thus make public to the participants from the university. Figure 2 presents the possible structure of prosumer market model at the university level, includes the rooftop solar panels which can be easily installed at the university buildings with battery system. Microturbine can be considered as a source, however, the availability of wind speed to produce an average power production is required, and necessary installation outside university compound is optional. Energy storage system can be used to maintain the system and backup even in the blackout or transition period. Electric vehicle station has to be one of the important sources to include in the prosumer consortium model though the accessible of EV will high depend on the location, nevertheless, as a dispersed storage system the injection of energy to the grid can be a benefit to the EV's owner. Usually, local consumer of university includes academic building, offices, residential units, lighting system and outreach buildings.

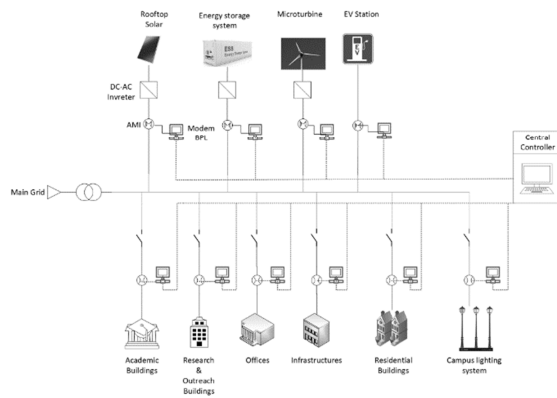

Figure 2 Feasible university grid system with a centralized controller.

Power line trading for prosumer and consumer nodes is assorted into substation level trading and centralized trading. Substation level trading depends on the total participants at each substation and one substation acts as one node of prosumer or consumer level. However, this research intends to emphasis centralized trading which all the nodes are connected at one communication line to central controller. Trading means good is a transfer between parties with a certain return with currency, currency shall be used on specified value of energy unit. Virtual currency will be a preferable system while performing the trading market. The central controller has an authority to has an interaction with a bank in order to inspect the purchasing of token by participants. The purchased token will be stored in the prosumer and consumer nodes.

During trading, the accessible power generation will be open to participants via website. The buyer has to make sure that has enough token to purchase the energy. When a buying request is made through the central control, it will make a decision based on energy production and token. Once the approval is passed, the token will be transferred to the destined seller account. Then the requested energy will be sent to the buyer node. The value of token on energy assets will be decided due to the availability of DERs and local participants. At some cases, a certain fee has to be paid to the central control as a transaction fee but not considered in this study. The decision of other trading events will be made by centralized controlled and it has the right to resolve the violation of seller and buyer agreements. General idea of trading is explained in Figure 3.

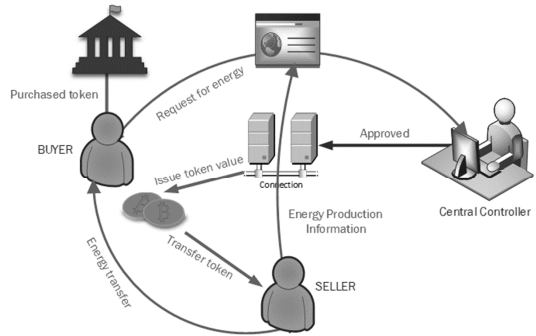

Figure 3 Actions of energy trading.

IV. CONCLUSION AND DISCUSSION

In this paper, the prosumer consortium market model for a university energy trading is presented. Many studies have been developed a community microgrid to create energy transferable mechanism and it has the nearly same architecture as this model. However, this proposed model can be tested in simulation case and then will be conducted in more research to be implemented. It has barriers to replace and upgrade an existing grid system and requires rigid ICT infrastructure in order to make strong and durable security. The use of hashing, cryptocurrency and encryption technology will be an optional method to solve trading and unauthorized access. A strong and transparent contract has to be made in order to accomplish integrity of the central controller and virtual account of each node must be protected from hacking and intrusion.

Advanced algorithms to make the prediction on renewable energy generation on prosumer or generation by hourly ahead or day ahead will be valuable information to purchase from consumers or buyers. Prediction of local demand is also an important factor to balance the local demand-supply. There is another research by authors that has been made a simulation in order to demonstrate the energy trading between prosumer nodes with virtual currency under the prosumer consortium market model. The local market model can provide financial benefit to the university and it can be assumed that can improve the understanding of energy market behavior to the learners from the university.

ACKNOWLEDGEMENTS

I would like to express my gratitude to Prof. Weerakorn Onngsakul for the guidance. Moreover, Dr. Jai Govind Singh, Dr. Chutiporn Anutrariya and Dr. Nimal Madhu Manjiparambil for their advice and encouragement for this research work. I would like to extend appreciation to Dr. Hien Vu Duc, supervisor of the AIT Energy laboratory for providing me the data.